\documentclass[twocolumn,prb,amsfonts,amsmath,amssymb,floatfix]{revtex4} 
\usepackage{color}
\usepackage{hhline}
\usepackage{mathrsfs}
\usepackage{graphicx}
\usepackage{dcolumn}
\usepackage{bm}
\usepackage{multirow}
\usepackage{booktabs}
\usepackage{afterpage}
\usepackage{amsmath}
\usepackage{ulem}

\begin{document}

\title{Quantifying the stability of the anion ordering in SrVO$_2$H}

\author{Masayuki Ochi}
\author{Kazuhiko Kuroki}
\affiliation{Department of Physics, Osaka University, Machikaneyama-cho, Toyonaka, Osaka 560-0043, Japan}

\date{\today}
\begin{abstract}
We investigate a text-book mixed-anion compound SrVO$_2$H using first-principles calculation,
to theoretically pin down the factors that stabilize its anion ordering.
We find that the {\it trans} preference by the characteristic crystal field in the VO$_4$H$_2$ octahedron
in addition to a coherent shrinkage along the V-H-V direction, taking place when such direction is consistent among neighboring hydrogens,
stabilize the anion ordering observed in experiment.
Our study gives an important clue for controlling the anion ordering in transition metal oxyhydrides.
\end{abstract}

\maketitle

\section{Introduction}

Transition metal oxides are of particular importance in condensed matter physics~\cite{SC}
due to their wide variety of physical properties.
Transition metal cations therein play a dominant role in determining their physical properties;
anions, on the other hand, have been regarded as subsidiary constituents.
However, {\it mixed-anion} strategy -- to say, employing multiple anions for controlling materials properties -- have recently attracted increasing attention~\cite{mixed_review}.
In particular, transition metal oxyhydride
is a special class of mixed-anion compounds, a unique electronic structure of which originates from several differences between O$^{2-}$ and H$^{-}$ ions.
Their different valence numbers enable the carrier control and lead us to the unexplored phases of iron-based superconductors~\cite{FeAs1,FeAs2}.
A unique crystal field for cation $d$ orbitals is realized by the 
inequivalency between the two anions~\cite{SVOH_Angew,SVOH_epitaxial,SVOH_JACS,SVOH_Yamamoto,SVOH_theory,SVOH_theory_wan,cRPAochi}.
Their different compressibility results in the strongly anisotropic deformation by external pressure for SrVO$_2$H~\cite{SVOH_Yamamoto}.
Also, their different valence orbitals, $p$ for oxygen and $s$ for hydrogen, drastically changes the dimensionality of the electronic structure in early transition metal oxyhydrides
because the $t_{2g}$ orbitals of cation cannot form a chemical bond with the H-$s$ orbital unlike with the O-$p$ orbitals, due to their different spatial symmetry~\cite{SVOH_Angew,SVOH_epitaxial,SVOH_JACS,SVOH_Yamamoto,SVOH_theory,SVOH_theory_wan,cRPAochi}.
Some recent theoretical studies focus on a unique electronic structure realized in transition metal oxyhydrides in the context of nickelate superconductivity~\cite{Held,Kitamine}.

\begin{figure}
\begin{center}
\includegraphics[width=8.4 cm]{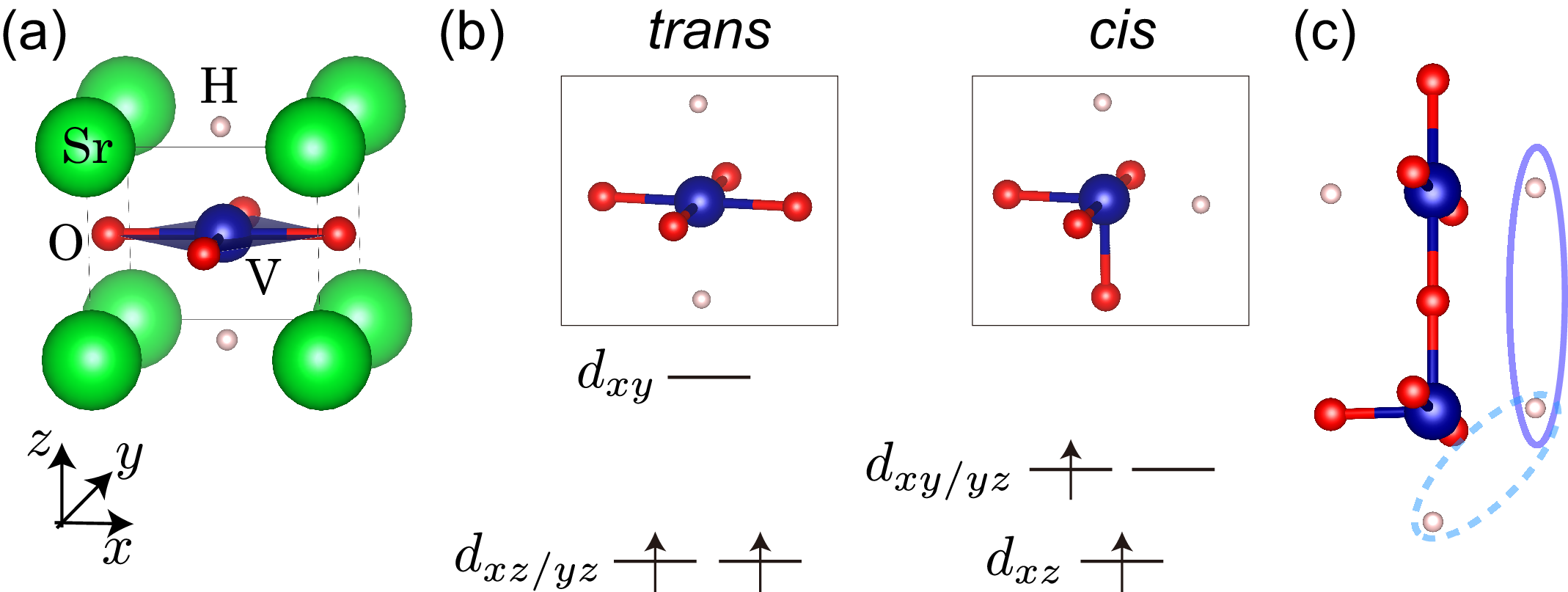}
\caption{(a) Crystal structure of SrVO$_2$H, (b) {\it trans} and {\it cis} configurations of hydrogens in VO$_4$H$_2$ octahedra shown together with their crystal fields for the V-$t_{2g}$ orbitals with $d^2$ filling,
(c) definition of a pair of two hydrogens, the number of which are counted as $n_{neighbor}$,
shown with a solid circle, together with a pair of hydrogens with their distance being $a/\sqrt{2}$ shown with a dotted circle (see details in the main text).
The green, blue, red, and white spheres represent Sr, V, O, and H atoms, respectively.
Although $t_{2g}$ is an inappropriate name for oxyhydrides with a lowered crystal-field symmetry in the strict sense of the term, we call the $d_{xy,yz,xz}$ orbitals the `$t_{2g}$ orbitals' for simplicity. The crystal structures shown in this paper were depicted using the VESTA software~\cite{VESTA}.}
\label{fig:1}
\end{center}
\end{figure}

Multiple anions, however, also bring an inevitable but crucial difficulty in treating a huge variety of anion configurations.
For molecules, a stable configuration of different kinds of ligands and its origin were investigated in many systems (e.g.,~Refs.~\onlinecite{mol1,mol2}).
For mixed-anion solids, however, the number of possible anion configurations exponentially increases by increasing the size of the simulation cell, which makes its theoretical analysis very difficult.
In usual cases, the anion configuration in mixed-anion solids observed in experiment is interpreted on the basis of the knowledge of molecular systems.
For example, in perovskite oxynitrides $AM$O$_2$N ($A$: alkaline earth metal, $M$: transition metal),
an energy profit by strong covalent bonding of $M$ $d_{\pi}$-N $p_{\pi}$ is considered to be maximized in a {\it cis}-$M$O$_4$N$_2$ octahedron rather than {\it trans} one,
on the basis of the knowledge of complex chemistry~\cite{mol1,mol2},
and thus is regarded to be the origin of the {\it cis}-preference in this system~\cite{oxynit_order}.
As another example, the stability of the anion ordering in Sr$_{n+1}$V$_n$O$_{2n+1}$H$_n$ ($n=1,2,\infty$)~\cite{SVOH_Angew,SVOH_JACS}, (Fig.~\ref{fig:1}(a) for the $n=\infty$ case, SrVO$_2$H) is usually explained by the {\it trans} preference of the $d^2$ filling in the isolated VO$_4$H$_2$ octahedron (e.g., Ref.~\onlinecite{SVOH_Angew}), as shown in Fig.~\ref{fig:1}(b).
These explanations are tested by some first-principles calculations~\cite{Fang,Wolff,SVTiOH},
but the number of configurations investigated in their studies is very limited, owing to the theoretical difficulty mentioned above.
Thus, it is indispensable to perform first-principles calculation for more extensive set of anion configurations.
In fact, such a calculation has recently become possible~\cite{BaVOH,Kaneko,BaScHO}.
Firm physical insight on the anion ordering is of great help to control anion ordering in experiments, which is possible only for very few cases~\cite{strain_control,kageyama_svoh_strain} in the present day.

In this paper, we investigate the relation between the total energy and the anion configurations in SrVO$_2$H, a text-book mixed-anion compound, using the first-principles calculation,
in order to theoretically pin down the factors that stabilize the anion ordering in this compound.
We find that the {\it trans} preference by the characteristic crystal field in the VO$_4$H$_2$ octahedron is indeed important for realizing the anion ordering, but is not sufficient to identify the stable structure. 
In fact, we identify another important factor: whether the structure allows a coherent shrinkage along the V-H-V direction, which can be induced by the different size of the O$^{2-}$ and H$^{-}$ ions and yields a sizable stabilization of the system. This study gives an important clue for controlling the anion ordering in transition metal oxyhydrides.

This paper is organized as follows. Section~\ref{sec:detail} presents our computational model and calculation conditions. In Secs.~\ref{sec:result} A--C, we present our calculation results performed under several conditions and discuss what stabilizes the anion ordering in SrVO$_2$H.
Additional discussions that reinforce our view including a rough estimate of some minor effects not included in our simulation are shown in Secs.~\ref{sec:result} D--F.
Section~\ref{sec:sum} summarizes this study.

\section{Computational model and methods\label{sec:detail}}
 
 \subsection{Computational model}

As a computational model, we considered possible anion configurations in Sr$_8$V$_8$O$_{16}$H$_8$, by replacing one third of oxygens with hydrogens in the $2\times 2\times 2$ supercell of SrVO$_3$, and optimized their structures in the way described in Sec.~\ref{sec:comput_cond}.
Because of a formidable number of possible anion configurations even for our relatively small-sized simulation cell,
we restricted our analysis to the configurations satisfying that each vanadium atom has two nearest-neighboring hydrogen atoms.
This constraint is energetically natural and often applied for analysis of anion (dis)ordering in materials with $M$O$_4X_2$ octahedra ($X$: anion other than oxygen)~\cite{corr_disorder}.
By using this assumption, the number of possible (symmetrically-inequivalent) crystal structures becomes around 250.

We focused on the following two quantities to characterize anion configurations.
One is the number of {\it trans} configurations of hydrogens around vanadiums, $n_{trans}$, which is up to eight since our supercell includes eight vanadium atoms.
The other one is the number of a pair of two hydrogens with specific relative positions as shown in Fig.~\ref{fig:1}(c) with a solid circle, $n_{neighbor}$.
Note that the atoms used for defining $n_{neighbor}$ are two vanadiums and two hydrogens surrounded with the solid circle in Fig.~\ref{fig:1}(c), and the other atoms shown in this figure are irrelevant to the definition.
In other words, $n_{neighbor}$ is the number of the square-shaped relative position of these four atoms: V-V-H-H.
Here, we avoid double-counting pairs of the equivalent hydrogens due to the periodic boundary condition, and then, e,g., $n_{neighbor}$ is eight for the $2\times 2\times 2$ supercell constructed from the experimental structure of SrVO$_2$H shown in Fig.~\ref{fig:1}(a).

These two quantities were chosen by the following reason.
First, we can represent the hydrogen configuration in Sr$_8$V$_8$O$_{16}$H$_8$ by a vector ${\bm x}=\{ x_1, x_2, \dots, x_{24} \}$ where $x_i$ denotes whether the $i$-th oxygen site in Sr$_8$V$_8$O$_{24}$ (i.e., a $2\times 2\times 2$ supercell of SrVO$_3$) is replaced with hydrogen ($x_i=1$) or not ($x_i=0$).
One can also uniquely specify a hydrogen configuration by a set of the values of all the $k$-th monomial $x_{i_1} x_{i_2} \dots x_{i_k}$ up to 
8th order, where $k$-th monomial corresponds to the $k$-body correlation among hydrogen (e.g., $x_5 x_7 x_9=1$ means that hydrogens occupy the 5th, 7th, and 9th anion sites).
Thus, we expect the relation,
\begin{equation}
E ({\bm x})= E_0 + \sum_i c_i x_i + \sum_{i<j} c_{ij} x_i x_j + \dots,
\end{equation}
where $E({\bm x})$ is the total energy with the hydrogen configuration ${\bm x}$ and $E_0, c_i, c_{ij},\dots$ are unknown parameters (coefficients).
This is the way used in the popular cluster-expansion method~\cite{cluster}, where atomic species in multicomponent systems such as alloy are denoted by a discrete variable and the energy of the system is represented with polynomials of those variables.
There is no need to consider terms including $x_i^n$ ($n\geq 2$) since $x_i^n=x_i$.
Because a small supercell was used in this study under the periodic boundary condition, we should concentrate on short-range correlations represented with a low-degree monomial.
There is no first order monomial that can be used to characterize anion configurations.
This is because the equivalency among all the oxygen sites in cubic SrVO$_3$ requires that not $x_i$ but $\sum_{i=1}^{24} x_i$ should be used to represent the total energy of this system (to say, all the coefficients $c_i$ in $E({\bm x})$ should be the same by symmetry), but $\sum_{i=1}^{24} x_i$ is constant (8) in our simulation.
By a similar discussion for the second order monomials, 
the number of equivalent hydrogen pairs $\sum_{\langle i j \rangle} x_i x_j$
can be used to represent the energy of this sytem, where the $i$-th and $j$-th hydrogens having specific relative positions that are equivalent by the crystal symmetry, are summed over.
Then, only the two quantities, $n_{trans}$ and $n_{neighbor}$,
are allowed if one considers monomials representing a pair occupation of hydrogens with their distance less than or equal to $a$, a lattice constant of cubic primitive cell.
Note that the inter-hydrogen length for the initial cubic structure is used in explanation just for simplicity.
We also note that the number of hydrogen pairs with their distance being $a/\sqrt{2}$ as shown with a dotted circle in Fig.~\ref{fig:1}(c), is nothing but the number of {\it cis} configurations, and thus is uniquely determined by $n_{trans}$ because we restrict the anion configurations to those where 
all the vanadiums have two adjacent hydrogens.
For simplicity, we omit the higher-order monomials in the following analysis.
It is an interesting future task to perform more elaborated statistical analysis such as the multiple regression analysis using many factors omitted in this study, while it might require a larger supercell without the restriction introduced in our simulation.
In the latter analysis, we shall see that, even by our simplification, the total energies well correlate with the two quantities focused in this study with a clear physical interpretation.

\subsection{Conditions of first-principles calculations \label{sec:comput_cond}}

We performed the structural optimization using the density functional theory with the Perdew-Burke-Ernzerhof parametrization of the generalized gradient approximation (PBE-GGA)~\cite{PBE} and the projector augmented wave (PAW) method~\cite{paw} as implemented in the {\it Vienna ab initio simulation package} (VASP)~\cite{vasp1,vasp2,vasp3,vasp4}. 
Both cell deformation and optimization of the atomic coordinates were allowed unless noted.
The following orbitals for each element are treated as core electrons in the PAW potentials: [Ar]$3d^{10}$ for Sr, [Ne] for V, and [He] for O.
We started our calculation from the G-type antiferromagnetic configurations of the V-$d$ orbitals.
Here, we need to include the spin polarization to describe the $d^2$ occupation into the crystal field shown in Fig.~\ref{fig:1}(b), which was pointed out to be a key to realize the {\it trans} preference of this system in the previous studies as mentioned in the introduction.
We included the $+U$ correction~\cite{DFTU1,DFTU2} with $U-J=3$ eV for the V-$d$ orbitals, which is a typical value for the $3d$ orbitals and is also consistent with our first-principles estimation by constrained random-phase approximation~\cite{cRPAochi}.
Here, we used the simplified rotationally invariant approach to the DFT$+U$ method introduced by Dudarev {\it et al}.~\cite{DFTU2}, where the correction term only depends on $U-J$ rather than the values of $U$ and $J$ themselves.
Note that SrVO$_2$H becomes gapless in simple PBE-GGA calculation without the $+U$ correction even when the spin polarization is taken into account, owing to the well-known underestimation of the band gap in PBE-GGA, while the system is insulating in experiments~\cite{SVOH_Angew}.
Further discussion on our choice of the $U-J$ parameter is presented in Sec.~\ref{sec:totE}.
While our simple DFT$+U$ calculation cannot describe the electron correlation effects completely, we consider that our simulation captures important factors determining the anion ordering as we shall see: a large structural change by introduction of hydrogen and the different crystal fields between {\it trans} and {\it cis} configurations. This is to some extent guaranteed by the spin-polarized insulating state observed in experimental studies at low temperature, where such a static approximation of the electron-electron interaction is somewhat validated. Electron correlation effects beyond our treatment, e.g., renormalization of the size of the crystal-field splitting, is an interesting issue but requires high computational cost and then out of scope of this study.
The initial crystal structure was generated in the way described before with a small random displacement, around $0.01$ in the crystal coordinate of the supercell for each direction, added to break the symmetry before the structural optimization.
 The plane-wave cutoff energy of 550 eV and a $6\times6\times6$ $\bm{k}$-mesh were used. 
 We applied Gaussian smearing with a smearing width of 0.1 eV.
 Structural optimization was continued until the Hellmann-Feynman force on each atom becomes less than 0.01 eV\AA$^{-1}$ for each direction.
 The spin-orbit coupling was not taken into account throughout this paper.

\begin{figure}
\begin{center}
\includegraphics[width=8.4 cm]{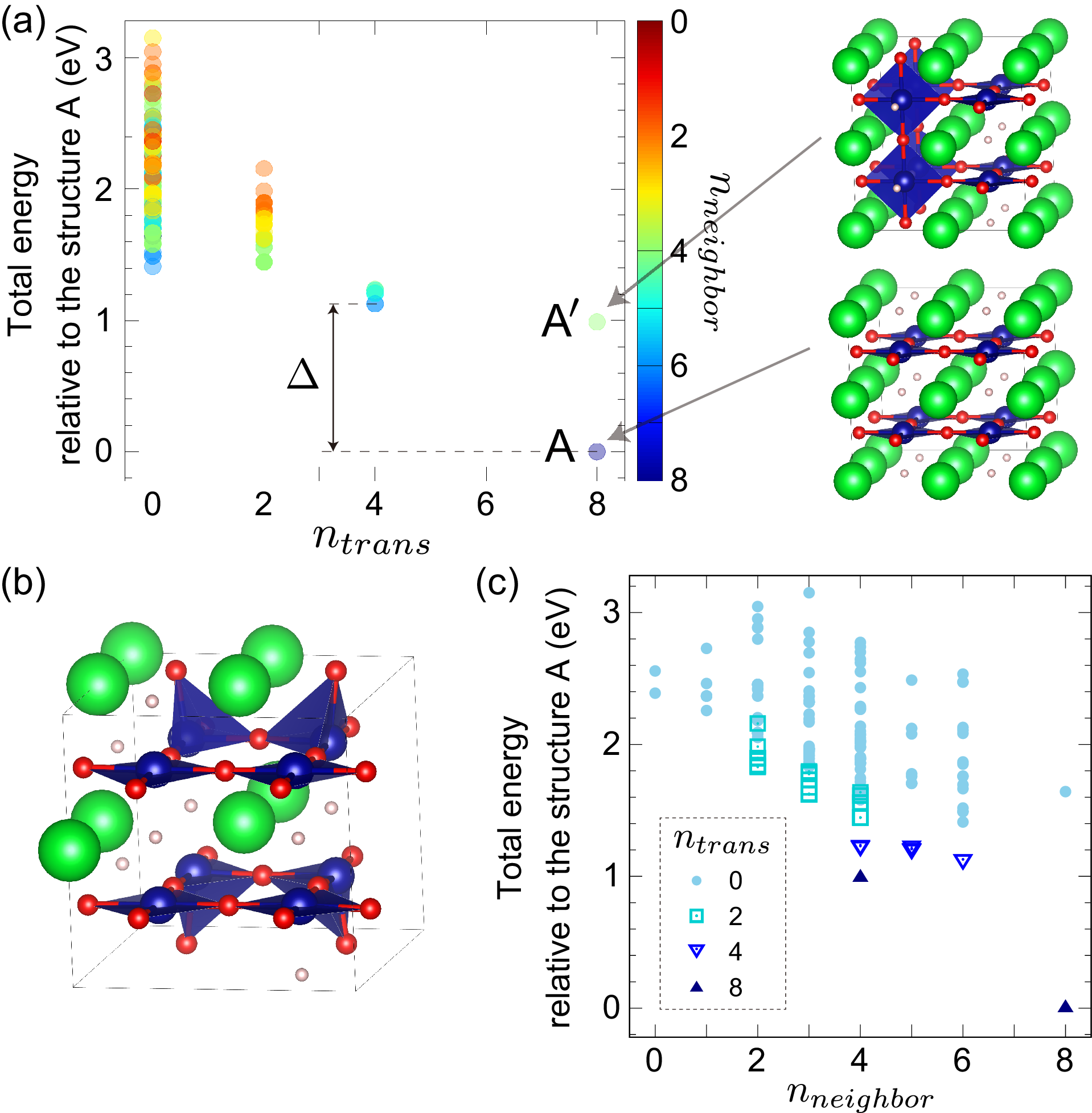}
\caption{(a) Total energy (eV) of all the calculated structures plotted against $n_{trans}$, where color of each point represents $n_{neighbor}$. The total energy for the $2\times 2\times 2$ supercell relative to that for structure A are shown throughout this paper. 
$\Delta$ defined in the main text, and the two {\it all-trans} structures called structures A and A$'$ in this paper, are also shown.
(b) Crystal structure B, which has the lowest energy except structures A and A$'$, and so used for calculating $\Delta$, in panel (a).
(c) Plot similar to panel (a) but shown against $n_{neighbor}$ instead of $n_{trans}$.}
\label{fig:2}
\end{center}
\end{figure}

\section{Results and discussions\label{sec:result}}

\subsection{Total energy and characteristic quantities \label{sec:totE}}

Figure~\ref{fig:2}(a) presents the total energies of computed structures plotted against $n_{trans}$ colored using a value of $n_{neighbor}$. 
Here, the two crystal structures with {\it all-trans} configurations called structures A and A$'$ in this paper are also shown.
We define $\Delta$ as the total energy of the most stable structure except A and A$'$, relative to that for structure A.
The corresponding crystal structure with $n_{trans}=4$, called structure B in this paper, is shown in Fig.~\ref{fig:2}(b).
Note that structure B is not used for calculating $\Delta$ under different conditions where another structure with $n_{trans}\neq 8$ is more stable than structure B, as shown in the following sections.
Because of the periodic boundary condition applied to our small supercell, $n_{trans}$ always takes an even value in our simulation. We find that structure A has the lowest energy in our simulation, which is consistent with experiments.
The total energies of the calculated configurations tend to be lowered by increasing $n_{trans}$.
However, the {\it all-trans} structure A$'$ has a sizably higher energy than that for structure A.
This means that the explanation of the anion ordering in the previous studies based on the stable $d^2$ crystal field for the {\it trans} configurations, is partially correct 
from the viewpoint of the negative correlation between the total energy and $n_{trans}$ as mentioned above
but insufficient to describe the stability of anion ordering in this system.
We shall come back to detailed analysis on the $n_{trans}$ dependence later in this paper, but first we focus on the difference between structures A and A$'$,
which should be a key to understand
what is missing in the description by $n_{trans}$.

One noticeable difference between structures A and A$'$
is whether hydrogens are placed onto the same ($ab$) plane or not, which can be captured by $n_{neighbor}$.
The colors of data points in Fig.~\ref{fig:2}(a) show that not only these two structures but also the other data points exhibit a tendency that a large $n_{neighbor}$ stabilizes the energy. In Fig.~\ref{fig:2}(c), the total energy is plotted against $n_{neighbor}$ instead of $n_{trans}$.
For each set of data points with the same $n_{trans}$, we can
verify the negative correlation between the total energy and $n_{neighbor}$.
A possible interpretation is as follows. It is known that the size of the H$^{-}$ and O$^{2-}$ ions are 
different and thus the lattice constants of SrVO$_2$H are quite anisotropic: $a=3.9290(6)$\AA\ and $c=3.6569(5)$\AA\  in experiment~\cite{SVOH_Angew}. In other words, introduction of hydrogen inclines the crystal to shrink along the V-H-V direction.
In structure A, the hydrogen occupation along the same direction allows the crystal to coherently shrink along the same direction ($c$ direction), while such coherent shrinkage is not allowed in structure A$'$. The coherent shrinkage by concomitant occupation of the neighboring hydrogen sites takes place when $n_{neighbor}$ is large.
We note that an existence of a hydrogen pair captured by $n_{neighbor}$
in the present small supercell means that hydrogens occupy anion sites in a row with an infinite length.
Thus, here we see that, not only a two-dimensional occupation of hydrogen in structure A, i.e., all oxygens in some $ab$ planes are fully replaced with hydrogens,
but also a one-dimensional occupation of hydrogen as captured by $n_{neighbor}$ effectively lowers the total energy.
While the small supercell used in this study is only capable of representing limited patterns of hydrogen configurations,
 we can partially see how the ordering of hydrogen occupation stabilizes the crystal structure.

Before moving on to further analysis, to check the robustness of our conclusion against the $U-J$ value, we calculated the total energy difference between structures A, A$'$, and B, using $U-J=5$ eV. For this purpose, we did not optimize the crystal structure but just used the optimized structure using $U-J=3$ eV. As a result, we find that the energy difference between structures A and B is $1.10$ eV for $U-J=5$ eV while it is $1.13$ eV for $U-J=3$ eV, and that between structures A and A$'$ is $0.97$ eV for $U-J=5$ eV while it is $0.99$ eV for $U-J=3$ eV. Therefore, we believe that our conclusion is less affected by the choice of the $U-J$ value in simulation.
 
\begin{figure}
\begin{center}
\includegraphics[width=8.4 cm]{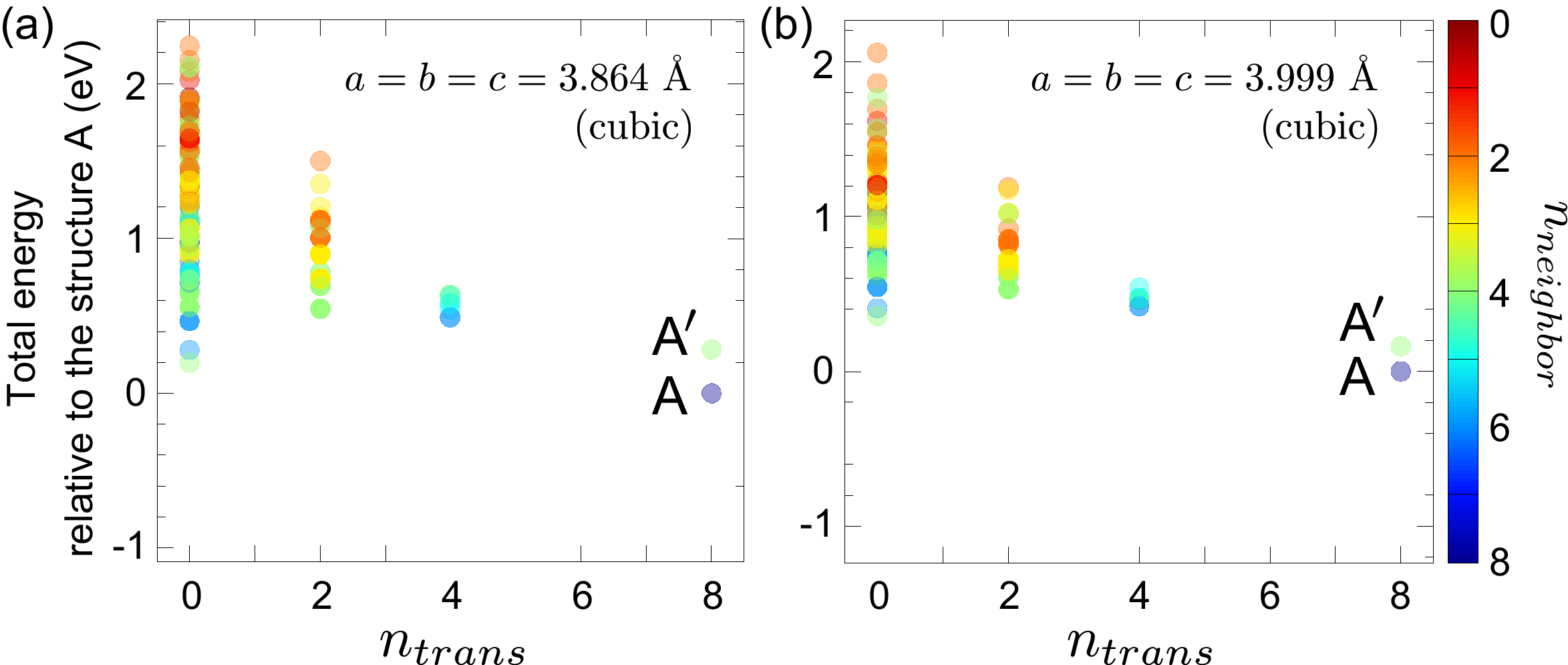}
\caption{(a) Total energy (eV) of all the calculated structures using the fixed lattice constant $a=b=c=$ 3.864 \AA \ for (a) and 3.999  \AA \ for (b), plotted in the same way as Fig.~\ref{fig:2}(a).}
\label{fig:3}
\end{center}
\end{figure}

\subsection{Effect of coherent shrinkage of V-H-V: insight from cubic-cell calculation}

To verify our view, we also calculated the total energies using a fixed cubic lattice, with $a=b=c=$ 3.864 \AA\ and 3.999 \AA\ for Figs.~\ref{fig:3}(a) and (b), respectively,
while atomic coordinates were relaxed.
Although the overall correlation between the total energies and $n_{trans}$, $n_{neighbor}$ are similar to Fig.~\ref{fig:1}(a), we find that the energy difference between structures A and A$'$ are drastically decreased: from $0.99$ eV in Fig.~\ref{fig:2}(a) to $0.29$ and $0.16$ eV in Figs.~\ref{fig:3}(a)--(b), respectively.
This is because the tetragonal deformation, i.e., $a=b\neq c$, is preferred for structure A but is not allowed in the cubic lattice.
It is interesting that first-principles calculation of unsynthesized KTiO$_2$H, although with restricted number of sampled structures, also pointed out that the different size between O$^{2-}$ and H${-}$ ions and a resulting shrinkage of the lattice along Ti-H-Ti are important in stabilizing structure A~\cite{Tsuneyuki}, where the crystal-field effect cannot take place for Ti-$d^0$  occupation therein.

Note that, structure A with the cubic lattice constants exhibits an octahedral rotation with the V-O-V angle of 168.5 and 170.1 degrees for Figs.~\ref{fig:3}(a) and (b), respectively, unlike the that for the relaxed tetragonal lattice shown in Fig.~\ref{fig:2}(a).
It is interesting that the lowest energy among the {\it all-cis} configurations with $n_{trans}=0$ relative to the total energy of structure A becomes much lower in Fig.~\ref{fig:3}(a) ($0.20$ eV) than in Fig.~\ref{fig:3}(b) ($0.36$ eV).
A definitive conclusion cannot be reached within our limited calculation data, but it seems that a smaller ionic radius of H$^{-}$ alleviates a cramped environment for atoms, and then stabilizes the crystal structure with the {\it cis} configurations, while it is not so active for structure A with VO$_2$ planes without hydrogen.
In the experimental study on SrCrO$_2$H, it was pointed out that the introduction of hydrogen into SrCrO$_3$ makes its tolerance factor closer to unity and then raises the N{\'e}el temperature~\cite{CrOH}.
It is an interesting future issue to investigate the effect of the octahedral rotation and tilt onto the stability of anion configuration, requiring some quantities to characterize such kinds of local distortion.
Here we just point out that this issue has some relevance to the coherent shrinkage captured by $n_{neighbor}$ dependence of the total energy and by the energy difference between structures A and A$'$: both of them originate from the different anion size and the resultant structural change.

\begin{figure}
\begin{center}
\includegraphics[width=8.4 cm]{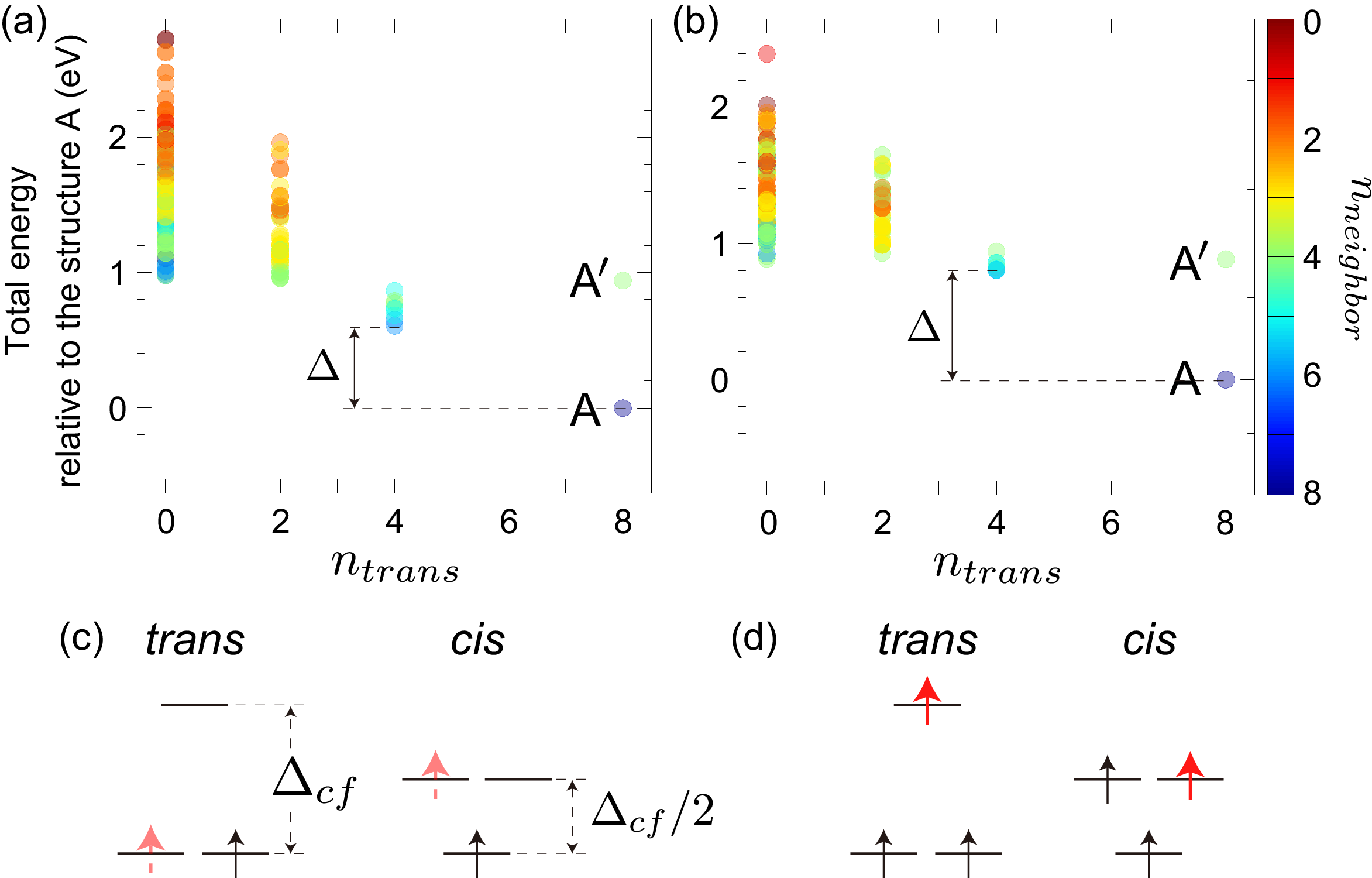}
\caption{(a) Total energy (eV) of all the calculated structures using the number of electrons (a) decreased or (b) increased by two, from the nominal Sr$_8$V$_8$O$_{16}$H$_8$, plotted in the same way as Fig.~\ref{fig:2}(a). (c) Schematic picture representing the $d^1$ filling into the {\it trans} and {\it cis} crystal fields, where removed electrons from Fig.~\ref{fig:1}(b) are shown by red broken arrows. (d) That for the $d^3$ filling where additional electrons from Fig.~\ref{fig:1}(b) are shown by red bold arrows.}
\label{fig:4}
\end{center}
\end{figure}

\subsection{Effect of the crystal-field splitting: calculations with different numbers of electrons}

To get insight into the $n_{trans}$ dependency, we performed the same simulation as Fig.~\ref{fig:2}(a) but with different numbers of electrons.
This is because the energy stabilization by the {\it trans} crystal field should be affected by a change of the number of electrons.
Figures~\ref{fig:4}(a) and (b) present the total energies calculated using the number of electrons increased or decreased by two in the supercell, respectively.
In other words, the electron occupation in the V-$d$ bands is $d^{1.75}$ and $d^{2.25}$ in average, for Figs.~\ref{fig:4}(a) and (b), respectively.
For these calculation, the same amount of the background positive charge was introduced.
We find that $\Delta$ is decreased from $1.13$ eV in Fig.~\ref{fig:2}(a) to $0.61$ and $0.81$ eV in Figs.~\ref{fig:4}(a)--(b), respectively.
The enhanced stability of the {\it trans} configuration in the $d^2$ occupation pointed out and tested
for some limited anion configurations in previous studies, is verified by our simulation for more intensive set of anion configurations.
In fact, experimental studies showed that hydrogens are randomly distributed in $A$TiO$_{3-x}$H$_x$ ($A=$ Ca, Sr, Ba, Eu)~\cite{TiOH1,TiOH2,TiOH3,TiOH4},
SrV$_{1-x}$Ti$_x$O$_{1.5}$H$_{1.5}$~\cite{SVTiOH}, 
and SrCrO$_2$H~\cite{CrOH}, where the number of $d$ electrons are increased or decreased from SrVO$_2$H.
Note that $\Delta$ can be decreased if one uses a larger supercell that can represent anion configurations not representable in our simulation.
In this sense, $\Delta$ estimated in our study is its upper bound, and so the stability of the {\it trans} configuration against the change of the electron number can be lost more easily than our simulation. 

Here we point out that a fewer electron occupation than $d^2$ can induce a Jahn-Teller instability in the {\it trans} crystal field that makes $d_{xz}$ and $d_{yz}$ inequivalent, and in fact it takes place in our simulation: V-O length becomes different by up to 0.09 \AA \ along the $a$ and $b$ directions in the optimized crystal structure.
It is interesting that a small reduction of electrons can provide an interesting platform of Jahn-Teller physics, while reducing a large number of electron will destabilize the crystal structures with the {\it trans} configuration.
We note that a Jahn-Teller instability can take place also for the $d^n$ ($1<n<3$) occupation in the {\it cis} configuration, but the crystal structure with {\it cis} configurations has a low symmetry in general even without the Jahn-Teller instability, and so it is difficult to quantify such an instability for the {\it cis} configurations.

\subsection{Alternative estimation of the crystal-field effect}

To verify our view on the crystal-field effect, we here make an alternative estimation of the energy stabilization by the crystal-field splitting.
In our previous study~\cite{cRPAochi}, we evaluated
the size of the crystal-field splitting between the $d_{xy}$ and $d_{xz/yz}$ orbitals, $\Delta_{cf} = 0.45$ eV, by constructing the Wannier functions~\cite{Wannier1,Wannier2}.
Suppose for simplicity that the energy levels of the $t_{2g}$ orbitals in the VO$_4$H$_2$ octahedron are simply determined by the number of hydrogens on the plane where the orbitals are extended (e.g., the $xy$ plane for the $d_{xy}$ orbital), 
the energy diagrams for the {\it trans} and {\it cis} configurations are obtained as shown in Fig.~\ref{fig:4}(c). Therefore,
the energy difference between the {\it trans} and {\it cis} configurations with $d^2$ occupation is around $\Delta_{cf}/2 = 0.23$ eV per vanadium.
This estimate corresponds to $-0.23$ $n_{trans}$ (eV) dependence of the total energy in the supercell, which amounts to the energy stabilization of $1.9$ eV for $n_{trans}=8$ compared with $n_{trans}=0$. This is roughly consistent with the energy separation among the data set of $n_{trans}=0$ and $n_{trans}=8$ at $n_{neighbor}=4$ and $n_{neighbor}=8$ in Fig.~\ref{fig:2}(c).
On the other hand, at $n_{neighbor}=4$ in Fig.~\ref{fig:2}(c), the energy separation between the data set of $n_{trans}=2$ and $n_{trans}=4$, and that between $n_{trans}=4$ and $n_{trans}=8$, are around 0.2 eV, which is a few times smaller than our rough estimate.
One possible reason is that, as we have seen, the introduction of hydrogen can induce a large structural change, which might alter the size of the crystal-field splitting from our rough estimate.
In addition, we can do a similar estimate of how $\Delta$ changes by adding or removing two electrons from or into the supercell.
By considering the {\it cis} and {\it trans} crystal fields shown in Figs.~\ref{fig:4}(c)--(d), the resultant change of $\Delta$ is $\Delta_{cf}=0.45$ eV for the both cases (i.e., making $d^1$ or $d^3$ filling for two vanadiums as we verified in the crystal structures used for calculating $\Delta$). 
This rough estimate is again consistent with our simulation for the change of $\Delta$, 
$0.52$ eV for Fig.~\ref{fig:4}(a) and $0.32$ eV for Fig.~\ref{fig:4}(b).
These observations, to some extent, validate our interpretation regarding the crystal field effect. Because the introduction of hydrogen can induce a large structural change, further investigation using more extensive data sets and characteristic quantities in a larger supercell would be an important issue to distinguish the crystal field effect and others more clearly, while the {\it trans}-preference is clearly verified in our simulation.

\subsection{Stabilization of random anion configuration by entropy effects}

Although both our simulation and experiment show that the anion-ordered structure A is the most stable, random anion distribution can be to some extent stabilized by entropy. Here, we provide a rough estimate on the effects of possible two kinds of entropy as follows.

First, the configuration entropy for fully random anion configuration is given by $S_{\mathrm{conf}} = -k_B \sum_i x_i \ln x_i$ per single anion site, where $x_i$ is the fraction of the anion $i$, i.e., $x_i=2/3$ for oxygen and $1/3$ for hydrogen in our case.
Thus, an energetic gain $-TS_{\mathrm{conf}}$ for fully random distribution compared with structure A with perfect anion ordering
is $-390$ meV for the $2\times 2\times 2$ supercell at $T=300$ K, which is much smaller than $\Delta$ in Fig.~\ref{fig:2}(a) (1.13 eV).
Note that the structures with random anion configurations have, in average, a larger energy by around 2 eV than structure A as shown in Fig.~\ref{fig:2}(a).

Second, the $d^2$ occupation in the {\it cis} configuration realized for random anion configuration can yield the electronic entropy.
To evaluate the upper bound of the electronic entropy effect, we assume that all the vanadiums have the {\it cis} configuration of hydrogens without a sufficiently large Jahn-Teller instability that breaks the degeneracy of the energy levels.
We also assume that the G-type antiferromagnetic ordering with a high-spin state $S=1$ is kept, 
where $d^2$ occupation is two-fold degenerate in the $t_{2g}$ manifold of the {\it cis} configuration (see Fig.~\ref{fig:4}(c)).
Then, the electronic entropy is given by $S_{el} = k_B \ln 2$ per vanadium, resulting in $-143$ meV stabilization for the $2\times 2\times 2$ supercell at $T=300$ K.
This is in the same order of magnitude but smaller than the stabilization by the configuration entropy described above.
While we assume the high-spin state ($S=1$) here, note that the low-spin state ($S=0$) rather stabilizes the {\it all-trans} configurations in terms of the electronic entropy effect. Additional entropy released for the paramagnetic state is not considered here because this is also acquired for the {\it all-trans} configurations above the transition temperature.

By considering the two types of entropies here, we can conclude that the entropy effects can be regarded as a relatively small correction to the total energy in SrVO$_2$H, although this effect can be important when $\Delta$ is small, e.g., in the case of different transition metal species and/or at high temperatures.

\subsection{Zero-point vibrational energy of hydrogen}

It is often the case that a relatively large zero-point vibrational energy due to a light mass of hydrogen can affect the stability of the crystal structure including hydrogen.
To check how the zero-point vibrational energy of hydrogens affects our conclusion, we performed phonon calculation for structures A and B because these two structures determine the value of $\Delta$ for Figs.~\ref{fig:2}(a) and Figs.~\ref{fig:4}(a)--(b), while the lowest-energy structure with $n_{trans}\neq 8$ is different from structure B in Fig.~\ref{fig:3}.
For simplicity, we calculated the $\Gamma$-point phonon frequencies to give a rough estimate of the zero-point vibrational energy: $\sum_{i} \hbar \omega_i/2$, where $i$ and $\omega_i$ denote the index of the phonon mode at the $\Gamma$ point and its frequency, respectively. For this purpose, we employed the finite displacement method as implemented in the \textsc{Phonopy}~\cite{phonopy} software in combination with VASP.
Other computational conditions are the same as those for the structural optimization presented in Sec.~\ref{sec:comput_cond}.

\begin{figure}
 \begin{center}
  \includegraphics[width=6.5cm]{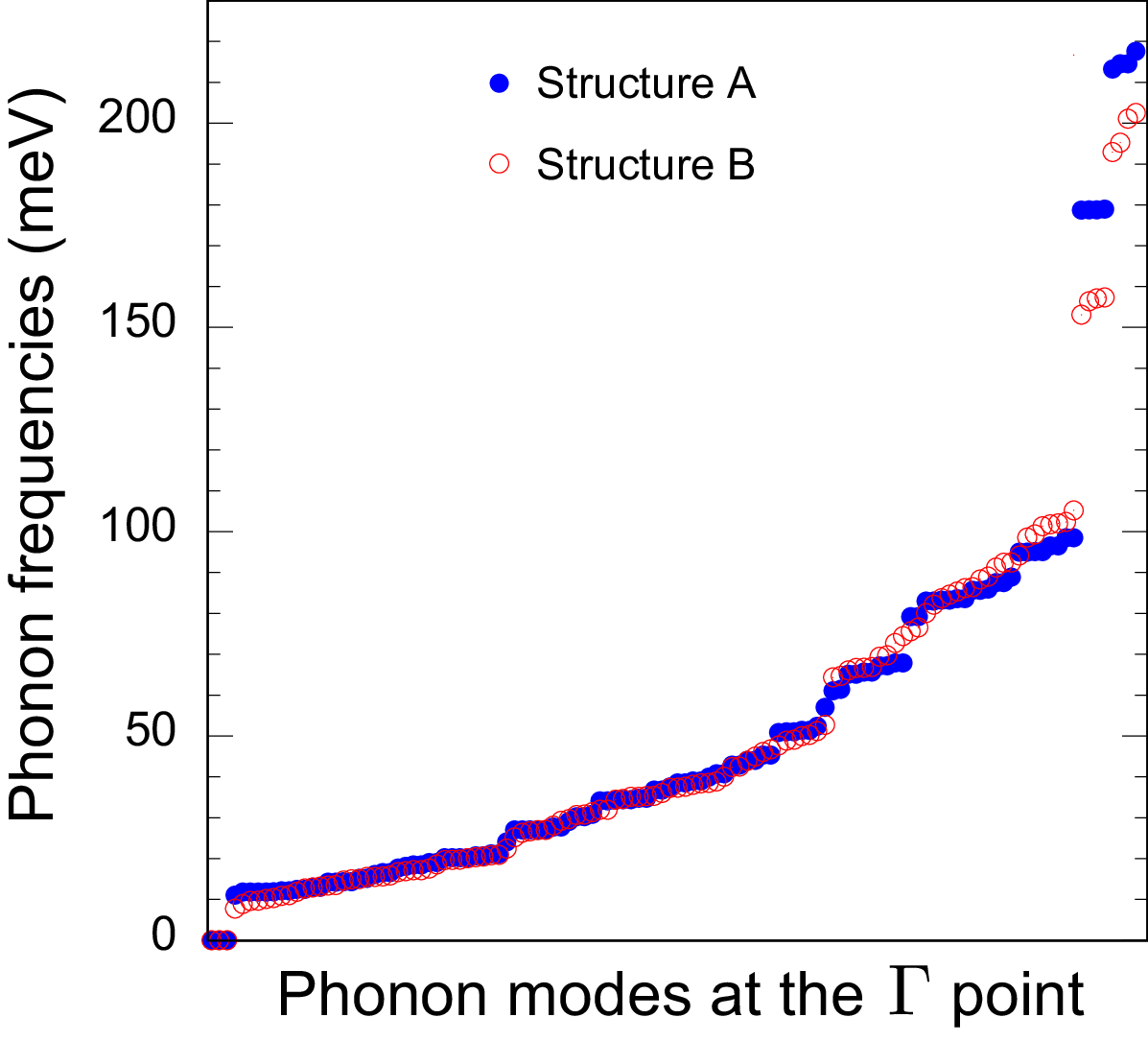}
  \caption{Phonon frequencies at the $\Gamma$ point for structures A and B. \label{fig5}}
 \end{center}
\end{figure}

Figure~\ref{fig5} presents the calculated phonon frequencies at the $\Gamma$ point for structures A and B. While the phonon frequencies are similar as a whole between the two structures, we find that the eight highest frequencies exhibit a sizable change of around 20 meV. As a result, the zero-point vibrational energy, $\sum_{i} \hbar \omega_i/2$, for these eight phonon modes in structure B is smaller by 79.6 meV than that for structure A. Because these high-frequency phonon modes represent the hydrogen moving toward the V-H-V direction, such an increase of the phonon frequency should take place when the V-H-V length becomes shorter.
Therefore, the energy stabilization by the shrinkage of the V-H-V length that can be captured by $n_{neighbor}$, is partially canceled by the increase of the zero-point vibrational energy.
However, the effect of the zero-point vibrational energy seems to be smaller than the energy profits discussed so far, hence does not affect our conclusion.

\section{Conclusion\label{sec:sum}}

We have investigated how the anion ordering is realized in SrVO$_2$H using the first-principles evaluation of the total energies for the structures with different anion configurations.
We have found that there are two important factors that stabilize the experimental crystal structures: 
one is the {\it trans} preference by the characteristic crystal field splitting in the VO$_4$H$_2$ octahedron
and the other one is the coherent shrinkage along the V-H-V direction.
The importance of the latter is demonstrated by the large energy difference between the {\it all-trans} structures A and A$'$.
In other words, structure A observed in experiment is stabilized by the {\it trans} preference at each VO$_4$H$_2$ octahedron together with the effective inter-octahedron interaction that favors the parallel alignment of the V-H directions.
Our study offers an important knowledge to control the anion ordering, which is indispensable for extracting desirable functionalities from the mixed-anion materials,
and thus expands the possibility of materials design.

\section{Acknowledgments}

Part of the numerical calculations were performed using the large-scale computer systems provided by the following institutions:
the supercomputer center of the Institute for Solid State Physics, the University of Tokyo, 
the Information Technology Center, the University of Tokyo, and the Cybermedia Center, Osaka University.
Computational resource from the Cybermedia Center was provided through the HPCI System Research Project (Project ID hp190022 and hp200007). This study was supported by JSPS KAKENHI (Grants No.~JP19H04697 and No.~JP19H05058) and JST CREST (Grant No.~JPMJCR20Q4).

\end{document}